\documentclass[prd,floatfix, nofootinbib, preprint]{revtex4-1}		%
\usepackage{amsmath,amssymb, bm, natbib}
\usepackage{epsfig}
\usepackage{graphicx}
\usepackage{color}

\newcommand{\RNum}[1]{\uppercase\expandafter{\romannumeral #1\relax}}

\begin{document}
\title{Non-leptonic decays of Charmed mesons into two Pseudoscalars}
\author{Aritra Biswas, Nita Sinha}
\affiliation{The Institute of Mathematical Sciences, C.I.T Campus, Tharamani, Chennai 600 113, India}

\author{Gauhar Abbas}
\affiliation{IFIC, Universitat de Val\`encia -- CSIC, Apt. Correus 22085, 
E-46071 Val\`encia, Spain}

\begin{abstract}
We examine the role of resonant coupled channel final state interactions (FSI), as well as weak annihilation and exchange contributions in explaining all the two body hadronic $D\rightarrow PP$ decay modes data. In the 
un-unitarized amplitudes we include modified Wilson coefficients with non-factorizable corrections as parameters. For the hadronic form factors, the z-series expansion method is used to get the $q^2$ dependence. 
The FSI effects are incorporated via a phenomenological approach with widths of resonances to various channels taken from observations where available, and others as additional parameters 
to be determined from fits of all the theoretical rates to the measured ones. Our results for the rather hard to explain $D^0\rightarrow K^+K^-, \pi^+\pi^- $ are in agreement with measured values. We demonstrate that 
both weak exchange as well as FSI effects are required to get the correct branching ratio for the $D^0\rightarrow K^0\bar{K^0}$ mode. Using our unitarized amplitudes we evaluate the strong phase 
difference between the amplitudes for $D^0\rightarrow K^-\pi^+$ and $D^0\rightarrow K^+\pi^-$ and find it to be in complete agreement with the recent BES III result.
\end{abstract}

\maketitle

\section{Introduction}

With the advent of the B factories, along with even the Tevatron having sufficient luminosities to perform excellent heavy flavour studies and more recently with the data pouring in from LHCb and BESIII, 
quark flavour physics is changing its role from being the ground for confirming the Kobayashi-Maskawa mechanism for $CP$ violation, to providing the observables that can test or constrain physics beyond the 
Standard Model. With improved precision measurements the goals of hadronic weak decays will cease to be, from seeking observables free from hadronic uncertainties, to actually precisely understanding the 
QCD effects.

Along with the plethora of data on semi-leptonic, hadronic, rare decays, $CP$ asymmetries and mixing in $B$ decays, discovery of charm mixing and hints of $CP$ violation in the charm sector, resulted in 
non-leptonic charm meson decays being a focus of attention, in the last few years~\cite{Artuso:2008vf,Ryd:2009uf}. Since $CP$ violation in charm, direct as well as in mixing, is expected to be negligible, 
any hint of $CP$ violation in charmed mesons is expected to be from physics beyond the standard model and hence charm may be instrumental in providing insights into new 
physics~\cite{Burdman:2003rs,Burdman:2001tf,Petrov:2010gy,D'Ambrosio:2001wg,Yabsley:2003rn,Close:2002fa,Fajfer:2001sa,Grossman:2006jg,Sinha:2007zz}. 
In fact, earlier FOCUS~\cite{Link:2000cu}, CLEO~\cite{Brandenburg:2001ze} and various other collaborations had produced many interesting results in the charm sector. 
The $3.2\sigma$ hint~\cite{Aaij:2011in,Collaboration:2012qw} of a difference of $CP$ asymmetries between the singly suppressed  $K^+K^-$ and $\pi^+\pi^-$ charmed decay modes resulted in a large volume of work
~\cite{Feldmann:2012js,Brod:2012ud,Brod:2011re,Cheng:2012wr,Bhattacharya:2012ah,Franco:2012ck,Isidori:2011qw,Grossman:2012ry,Atwood:2012ac,Giudice:2012qq,Hiller:2012wf,
Hochberg:2011ru,Altmannshofer:2012ur,Isidori:2012yx,Hiller:2012xm,Buccella:2013tya,Muller:2015lua}, mostly using different models of New Physics, to explain the result.
The hint has since then been slowly moving towards zero and currently there seems to be no evidence for any direct $CP$ violation in charm in any mode~\cite{Charles:2014dga}. Even though this hint for 
the $CP$ asymmetry slowly disappeared, from all the recent work done on the charm decays to two pesudoscalar modes motivated by this hint, it was clear that it is critical to first understand the observed 
branching ratios of all the charmed hadronic decay modes well, within the Standard Model, before any observation of an anomalous rate or any new CP asymmetry can be claimed to be, due to the presence of New 
Physics. 

However, this is not an easy task. The mass of the charm quark ($1.275$ GeV) makes it very difficult to come up with a proper theoretical technique for calculation of hadronic charmed meson decays. 
The charm, unlike the bottom quark, is not sufficiently heavy to allow realization of the infinitely heavy quark limit. Therefore, the well-known theoretical approaches based on QCD, for example, heavy quark 
effective theory~\cite{Politzer:1988bs,georgi1991heavy}, QCD factorization~\cite{Beneke:2000ry,Beneke:1999br}, the perturbative QCD approach~\cite{Keum:2000ph,Keum:2000wi,Lu:2000em,Lu:2000hj} and the 
soft-collinear effective theory~\cite{Bauer:2001yt}, 
which lead to very satisfactory predictions for $B$ decays, cannot be used to explain data in the case of charmed mesons. 
Furthermore, the charm quark is also not light enough for a chiral expansion to be applicable.

In the absence of any other reliable and effective theoretical methods, the factorization approach is still one of the most successful ways to study two-body charm meson 
decays~\cite{Bauer:1986bm,Wirbel:1985ji}. However, it is well known now that in the naive factorization approach, calculation of Wilson coefficients of effective operators faces the problem of 
$\gamma_5$- and renormalization scheme dependence. These difficulties can be overcome in the frame-work of the $\textquoteleft$generalized factorization approach' where Wilson coefficients are effective 
and include important non-factorizable (NF) corrections~\cite{Ali:1998gb,Ali:1998eb}. 

In the past there was another attempt to explain hadronic D decays using the so called large $1/N_c$ (where $N_c$ is number of colour degrees of freedom) approach~\cite{Buras:1985xv}. It was observed that 
dropping Fierz transformed terms characterized by $1/N_c$ can narrow the gap between predictions and observations up to a satisfactory level. The calculations based on QCD sum rules 
showed that Fierz terms were certainly compensated by the NF corrections~\cite{Blok:1986um,Blok:1986hm,Blok:1986sn}.   

There exists another model independent so called $\textquoteleft$quark diagram' or $\textquoteleft$topological diagram' approach in the 
literature~\cite{PhysRevD.36.137,Chau1989285,Rosner:1999xd,Chiang:2001av,Chiang:2002mr,Bhattacharya:2008ss,Bhattacharya:2009ps,Bhattacharya:2010uy,Cheng:2010ry} where all two-body non-leptonic weak 
decays of heavy mesons are expressed in terms of distinct quark diagrams, depending on the topologies of weak interactions, including all strong interaction effects. It is based on $SU(3)$ symmetry, and 
allows extraction of the quark diagram amplitudes by fitting against experimental data. 
However SU(3) breaking effects in charmed meson decays have been shown to be important and need to be carefully incorporated~\cite{Chau:1993ec,Wu:2004ht} .

The importance of final state interactions (FSI) in nonleptonic charm decays had been realized and discussed in several papers~\cite{Lipkin:1980es,Donoghue:1979fp,Sorensen:1981vu,Kamal:1980mw} in 
the early 80's, where the authors had been intrigued by the anomalies in the observed branching ratios of the Cabibbo favoured (CF), neutral versus the charged $K\pi$ modes, the differing rates of the 
singly Cabibbo suppressed (SCS) $K^+K^-$ and $\pi^+\pi^-$ modes, followed by measurements of rates of few other modes that had unexpected suppression/enhancement. Many of these were conjectured to be due 
to FSI. Surprisingly, even in the last couple of years, in  many of the papers that worried about the charm CP asymmetry problem, these old puzzles were still considered unresolved.

Even for the case of  hadronic B meson decays, the role of FSI's is being examined rather carefully in the last few years~\cite{Chua:2005dt,Gronau:2012gs}. 
 The mass of the charmed meson lies right in the heart of the resonance region. Hence, resonant final state rescattering is bound to play a bigger role in the two body hadronic charm decays and needs 
to be evaluated. Of course  dynamical calculations of these long distance effects are not possible and hence they can only be determined phenomenologically after comparison of the theoretical estimates 
with experimental data. Unitarity constraints play an important role in providing the theoretical estimates.  

Another contribution in hadronic two body decays that has been debated for a long period is that of the weak annihilation and exchange diagrams. One of Rosen's proposal~\cite{Rosen:1979us} 
had been that the W exchange diagrams may be large and since this appears only in  $D^0$ and not in $D^+$ decays, it could account for the difference in the lifetimes of these two mesons. Bigi and 
Fukugita~\cite{Bigi:1979zv} had then proposed several $D$ and $B$ meson decay modes that could be the smoking gun signals of the W-exchange contributions and yet, when the mode 
$D^0\rightarrow \phi{\bar{K}}^0$ was observed, it was argued~\cite{Donoghue:1986nu} that it could have been generated from the decay mode $D^0\rightarrow K^*\eta$, with this final state rescattering to the  
$\phi{\bar{K}}^0$ mode. 
Annihilation type contributions along with FSI's were incorporated in the hadronic two body vector-pseudoscalar modes of charmed meson decays in Ref.~\cite{Kamal:1988ub}. Studies using 
the quark diagram approach of Ref.~\cite{Cheng:2002wu} had also indicated that annihilation type contributions are needed to explain the observed data.

	In this paper, we study the role of FSI in the two body $D$ ($D$ here can be any of the $D^0$, $D^+$, or $D_s^+$) meson decays. We assume that FSI effects are dominated by resonance states close to 
the mass of $D$ mesons. In fact, there exist isospin $0$, $1$ and $1/2$ resonances near the $D$ mass, that may contribute to rescatterings among different channels in these respective isospin states and 
enhance/suppress some of the decay rates. 
	In the next section, we give the formalism for the calculation of the un-unitarized amplitudes, using a modified factorization approach, where the effective Wilson coefficients include 
NF corrections. This is in analogy with the QCD factorization approach of Beneke-Neubert for hadronic $B$ meson decays~\cite{Beneke:2003zv}, where however, using the hard scattering approach, the 
NF corrections are calculable in heavy quark approximation. However, for charm, since this approximation fails, these NF corrections are not calculable and are left as parameters. 
We also indicate our parametrization of the annihilation contributions and discuss our inputs: the decay constants, and the form factors, for which we have used a z-series expansion approach. In 
Sec.~\ref{Final State Interactions}, we discuss the need to incorporate additional long distance FSI effects and show how this can be done with a K matrix formalism for coupled channels. 
Using the observed widths, masses and known decay rates 
of the resonances to the various channels to evaluate the diagonal elements and leaving the unknown elements of the K matrix as parameters, the unitarized amplitudes are calculated (as discussed 
in~\cite{Kamal:1988ub}) to estimate the branching ratios of all the SCS, CF and doubly Cabibbo suppressed (DCS) $D\rightarrow PP$ decay modes. In Sec.~\ref{Numerical Analysis}, we list the isospin 
decomposition of all the decay modes, the parameters that need to be determined from our fits as well as the errors in theoretical inputs used. We list all the branching ratios after our numerical 
$\chi^2$ fits as well as the values of the fitted parameters. Finally we conclude in Sec.~\ref{Conclusions}.

\section{The Un-unitarized Amplitudes}
\subsection{Weak Hamiltonian and Wilson coefficients}
\begin{figure}
  	\begin{center}
   		\includegraphics[scale=.75]{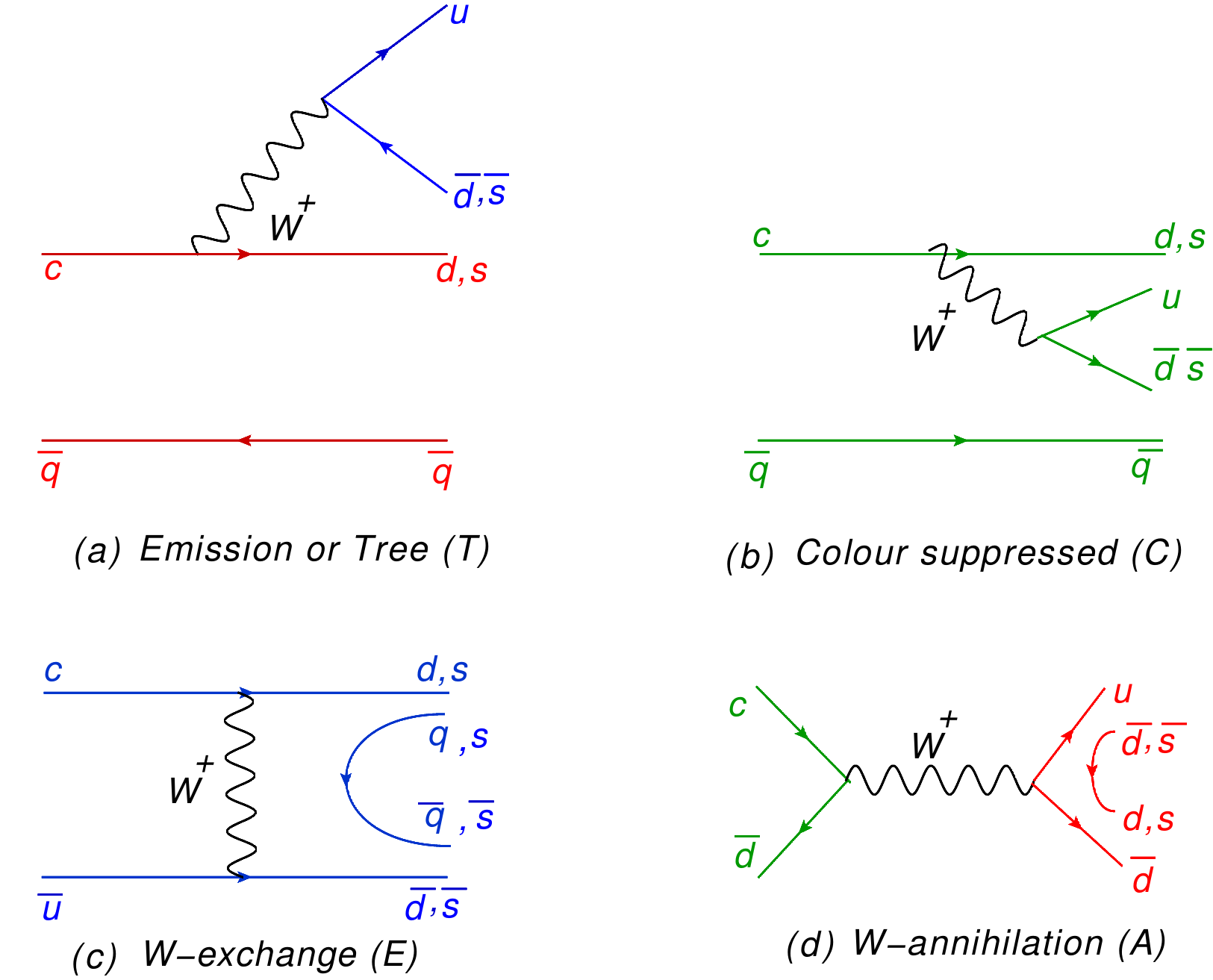}
                \caption{The dominant quark diagram amplitudes }
  		\label{Diagrams}
  	\end{center}
  \end{figure}

The study of weak decays of charmed mesons to two body hadronic modes necessarily requires a careful evaluation of the strong interaction corrections. The weak effective Hamiltonian may be expressed in 
terms of coefficient functions, which incorporate the strong interaction effects above the scale $\mu\sim m_c$ and the  current-current operators as:
\begin{eqnarray}
\mathcal{H}_{w}=\frac{G_F}{\sqrt{2}}[C_1(\mu)O_1(\mu)+C_2(\mu)O_2(\mu)]+h.c.~.
\label{(Hamiltonian)}
\end{eqnarray}
where, $G_F$ is the Fermi coupling constant, $C_1$ and $C_2$ are the Wilson coefficients and the operators are,
\begin{eqnarray*}
O_1&=&(\bar u_\alpha q_{2\alpha})_{V-A}(\bar q_{1\beta}c_\beta)_{V-A}\\
O_2&=&(\bar u_\alpha q_{2\beta})_{V-A}(\bar q_{1\beta}c_\alpha)_{V-A}~.
\end{eqnarray*}
 $\alpha$ and $\beta$ in the above are colour indices, while $q_1$, $q_2$ can be either the $d$ or the $s$ quark. The quark diagrams dominantly contributing to the branching ratios of 
$D\rightarrow P_1P_2$~\cite{Cheng:2010ry} are the colour-favoured Tree amplitude $T$, the Colour-suppressed amplitude $C$, the W-exchange amplitude $E$ and the W annihilation amplitude $A$, shown in 
Fig.~\ref{Diagrams}.

Penguin contributions in charmed meson decays are highly suppressed as the dominant down type quark contribution to the flavour changing neutral current $c\rightarrow u$ transition is from the $b$ quark 
which is accompanied by  
 the presence of the tiny product, $V_{cb}^*V_{ub}$ of the CKM matrix elements. Hence, the two operators in Eq.~(\ref{(Hamiltonian)}) are sufficient for calculating the amplitudes and branching ratios of 
the $D\rightarrow PP$ modes.  

In the naive factorization approach, the matrix element of the four-fermion operator in the heavy quark decay is replaced by a product of two currents. The amplitudes for the non-leptonic 2 body decay modes 
are then the product of a transition form factor and a decay constant. However, NF corrections must exist; while such corrections for scales larger than $\mu$ are taken into consideration in 
the effective weak Hamiltonian, 
those below this scale also need to be carefully incorporated. In the QCD factorization approach for $B$ meson decays~\cite{Beneke:2000ry,Beneke:2000fw,Beneke:1999br}, 
these NF corrections are handled using the hard scattering approach, where the vertex corrections and the hard spectator interactions are 
added at the next to leading order in $\alpha_s$ and its accuracy is limited only by the corrections to the heavy quark limit. 
But, in the case of charm decays, where the heavy quark expansion is not a very good approximation, it is 
best to parametrize these NF corrections and then determine them by fitting the theoretical branching ratios with the experimental data. In the diagrammatic approach of Ref.~\cite{Cheng:2010ry} 
also, either the Wilson coefficients themselves or the NF corrections appearing in the Wilson coefficients are determined from fits to data.

Hence, we write the scale dependent Wilson coefficients, modified to include the NF corrections which are parametrized by $\chi_1$ and $\chi_2$ with their respective phases $\phi_1$ and 
$\phi_2$ as, 
\begin{eqnarray}
a_1(\mu) & = & C_1(\mu) + C_2(\mu)\left(\frac{1}{N_c}+\alpha(\mu)\chi_1\mathrm{e}^{i\phi_1}\right)\\
a_2(\mu) & = & C_2(\mu) + C_1(\mu)\left(\frac{1}{N_c}+\alpha(\mu)\chi_2\mathrm{e}^{i\phi_2}\right)~.
\end{eqnarray} 
  The dominant Tree and Colour amplitudes for $D\rightarrow P_1P_2$, where $P_1$ is the final meson which carries the spectator quark, while $P_2$ represents the meson emitted from the weak vertex (as 
depicted in Fig.~\ref{Diagrams}), are then written as:                                             
  \begin{equation}
  T(C)=\frac{G_F}{\sqrt{2}}V_{CKM}a_1(\mu)(a_2(\mu))f_{P_2}(m_D^2-m_{P_1}^2)F_0^{DP_1}(m_{P_2}^2)~,\\
  \end{equation}
  where, $f_{P_2}$ is the $P_2$ meson decay constant and $F_0^{DP_1}(m_{P_2}^2)$ denotes the transition form factor for $D\rightarrow P_1$ evaluated at $m_{P_2}^2$. 
We follow the prescription of Ref.~\cite{Li:2012cfa} and choose the scale $\mu$ to be the energy release in individual decay processes rather than fixed at $m_c$. This scale which is dependent on the final 
state masses, allows for SU(3) breaking, additional to that coming from different decay constants and form factors. This scale is taken to be, $\mu= \sqrt{\Lambda m_D (1-r_2^2)}$, where, 
$r_2^2=m_{P_2}^2/m_D^2$ and $\Lambda$ is another free parameter. $\Lambda$, $\chi_1$, $\chi_2$, $\phi_1$ and $\phi_2$ are taken to be universal for all the decay modes and are fitted from experimental 
data.

\subsection{Weak Annihilation Contributions}
For a long time, W-exchange and W-annihilation contributions used to be neglected due to the so-called helicity suppression. However, observation of many decay modes of charmed and bottom mesons, which are 
possible only via the annihilation or exchange diagrams have indicated that these contributions could be substantial. These short distance weak annihilation effects were hence included in the diagrammatic 
approach.
In principle these could result from rescattering, even in the absence of annihilation and exchange processes. In fact, weak annihilation topologies were assumed to be induced by nearby resonances through 
FSI's in Ref.~\cite{Cheng:2002wu}. The authors in~\cite{Cheng:2002wu} as well as Ref.~\cite{Zenczykowski:1996xd}
	 use SU(3) to relate the couplings of the final state mesons with the resonances. This leads to the result that the long distance  W-exchange contribution can be induced by a tree amplitude, 
while W-annihilation can be induced by a color suppressed internal W-emission. The resonant FSI modify the W-exchange and W-annihilation amplitudes but the T and C amplitudes are unaffected. We emphasize 
that the assumption of SU(3), plays an important role in these results. In a most general coupled channel formalism, all contributions in the various channels will be affected by the resonant FSI, as will be 
shown in Sec \RNum{3}. We parametrize the W-exchange and W-Annihilation contributions in the  amplitudes by $\chi^{E(A)}$ 
and estimate them from phenomenological fits to data. The exchange(annihilation) amplitudes are hence written as:
\begin{equation}
E_{q,s}(A_{q,s}) = \frac{G_F}{\sqrt{2}}V_{CKM} C_1(\mu)(C_2(\mu)) \chi_{q,s}^{E(A)} \frac{C_F}{N_c^2}f_D f_{P_1}f_{P_2}~.
\end{equation}
 Since the initial charmed meson is annihilated and both the final mesons are produced from the  weak vertex, after the production of a quark-antiquark pair from a gluon, these amplitudes are a product of 
the decay constants of the initial $D$ meson ($f_D$) and that of $P_1(f_{P_1})$ and $P_2(f_{P_2})$. Apart from this, the strengths of the exchange(annihilation) amplitudes $\chi_{q,s}^{E(A)}$ are assumed to 
be the same for all modes and the subscripts distinguish the contributions of the pair production of the light quark-antiquark from that of the strange pair. Since the annihilation and exchange contributions 
are necessarily non-factorizable, they depend only on $C_{1,2}$ rather than the modified coefficients $a_{1,2}$. 
 
 We wish to emphasize that for the case of decays to two pseudoscalar mesons, it has been shown in Ref.~\cite{Beneke:2003zv} that in the annihilation contributions the quark-antiquark pair production happens 
with gluon emission from the initial state quark. Hence, this contribution is independent of the FSI effects that obviously involve the final state quarks and are discussed in 
Sec.~\ref{Final State Interactions}. Hence inclusion of both weak annihilation/exchange contributions as well as FSI will not amount to double counting, as pointed out in Ref.~\cite{Lai:2005bi}.   

 The scale of the Wilson coefficients for the exchange and annihilation amplitudes must depend on both the mass ratios, $r_{1,2}= m_{P_{1,2}}/m_D$ and is taken to be, 
$\mu= \sqrt{\Lambda m_D(1-r_1^2)(1-r_2^2)}$~. 

\subsection{Non-perturbative Inputs: Form Factors and Decay Constants}
\label{FF} 
 
We start by specifying our convention for the different mesons involved in our analysis:
\begin{eqnarray*}
 &\pi^+ = -u\bar{d},\;\;\;\;\;\; \pi^- = d\bar{u} ,\;\;\;\;\;\; \pi^0= \frac{u\bar{u}- d\bar{d}}{\sqrt{2}},& \\
 &K^0 =  d\bar{s} ,\;\;\;\;\;\; \bar{K^0} =  -s\bar{d},\;\;\;\;\;\; K^+= u\bar{s} ,\;\;\;\;\;\; K^-= s\bar{u},&\\
 &D^0=c\bar{u},\;\;\;\;\;\; D^+=-c\bar{d},\;\;\;\;\;\;D_s^+=c\bar{s}.&
\end{eqnarray*}

 In the $D\rightarrow P_1$ transitions, the matrix element of the vector current is written in terms of the form factors $F_+$ and $F_0$ as,
\begin{eqnarray}
\langle P_1(p^\prime)|\bar{q}\gamma^\mu c|D(p)\rangle\equiv F_+(q^2)(p^\mu + p^{\prime \mu}-\frac{m_D^2-m_{P_1}^2}{q^2}q^\mu)+F_0(q^2)\frac{m_D^2-m_{P_1}^2}{q^2}q^\mu,~
\label{(FF)}
\end{eqnarray}
where $q\equiv p-p^\prime$. The matrix element for the production of the second meson $P_2$, is given by,
\begin{equation}
 \langle P_2(q)|\bar{q_1}\gamma_\mu q_2|0\rangle = if_{P_2} q_\mu~.
 \label{(DecayConst.)}
\end{equation}
 Hence, in the amplitude of the non-leptonic two pseudoscalar decay modes of charmed mesons involving the product of the two matrix elements specified in Eqs.(6) and (7), only the transition form factor 
$F_0$ appears. Transition form factors can in principle be experimentally measured from the semi-leptonic decays, however, in the massless lepton limit, only the $F_+(q^2)$ contributes to the semi-leptonic 
amplitude distributions. However, the semi-leptonic information is still useful, since at zero momentum transfer the form factors obey the kinematic constraint $F_0(0)=F_+(0)$. 
The $q^2$ dependence of the  $F_0$ on the other hand is accessible only with massive leptons in the semileptonic decays or in lattice simulations.
Simple and modified pole models have been widely used to parametrize the $q^2$ dependence of the form factors, but these have poor convergence properties. Recently the z-expansion~\cite{Becher:2005bg,Hill:2006ub} has been 
introduced as a model independent parametrization of the $q^2$ dependence of form factors over the entire kinematic range and has been shown to have improved convergence properties. In this approach, based on analyticity 
and unitarity, the form factors are expressed as a series expansion in powers of $z^n$, where z is a non-linear function of $q^2$, with an overall mutiplicative function accounting for the sub-threshold 
poles and branch cuts,
\begin{equation}
F(t) = \frac{1}{P(t)\phi(t,t_0)}\sum_{\mathrm k=0}^{\inf} a_k(t_0)z(t,t_0)^k~.
\label{z-exp}
\end{equation}
The series coefficients and prefactors can only be determined from fits to lattice or experimental data. In fact, CLEO collaboration has determined these coefficients for the $D\rightarrow \pi,K,\eta$ form 
factors from the semileptonic decays but, in the massless lepton limit. Hence, for $F_0(q^2)$, we use lattice results to determine the first two coefficients.

In the Becirevic and Kaidalov (BK) ansatz~\cite{Becirevic:1999kt}, 
\begin{equation}
F_0(q^2) = \frac{F_0(0)}{1-\frac{q^2}{\beta m_D^{* 2}}}, \footnote{It has been pointed out in Ref.~\cite{Hill:2005ju} that $F_0$, but not  $F_+$, can be modeled by a single pole.}
\end{equation}
where, $m_D^*$ is the mass of the vector meson with flavour $c\bar{d}$ or $c\bar{s}$, depending on the transition being $c\rightarrow d$ or $c\rightarrow s$ respectively. $F_0(0)$ and $\beta$ are parameters 
to be fitted to experimental data and in fact, in Ref.~\cite{Bernard:2009ke}, the Fermilab and Lattice MILC collaborations, have fitted these parameters to CLEO-c data. The normalization $f(0)$ and shape 
parameter determine the physical observables describing the form factors at large recoil and are given by, 
\begin{equation}
f(0) \equiv F_+(0) = F_0(0),~~~~~\displaystyle \frac{1}{\beta}\equiv \frac{(M_H^2-M_L^2)}{F_+(0)} \frac{d F_0}{d q^2}{\huge |}_{q^2=0}~.
\end{equation}
Using these input parameters we can determine the first two coefficients of the series expansion in Eq.~(\ref{z-exp}) for $F_0(q^2)$.   

Few details regarding the z series form factor expansion can be found in Appendix B. 
If, the series is rapidly converging, even two coefficients may be sufficient to determine the $q^2$ dependent form factor $F_0(q^2)$.  
Equating the normalization and slope obtained using Eq.~(\ref{z-exp}) to that obtained from the lattice parameters ($f(0),\beta$), which in turn had been obtained by fits to experimental data, we can obtain 
the z-expansion series (up to linear order) for all the form factors.

\begin{figure}
  	\begin{center}
  		\includegraphics[scale=0.55]{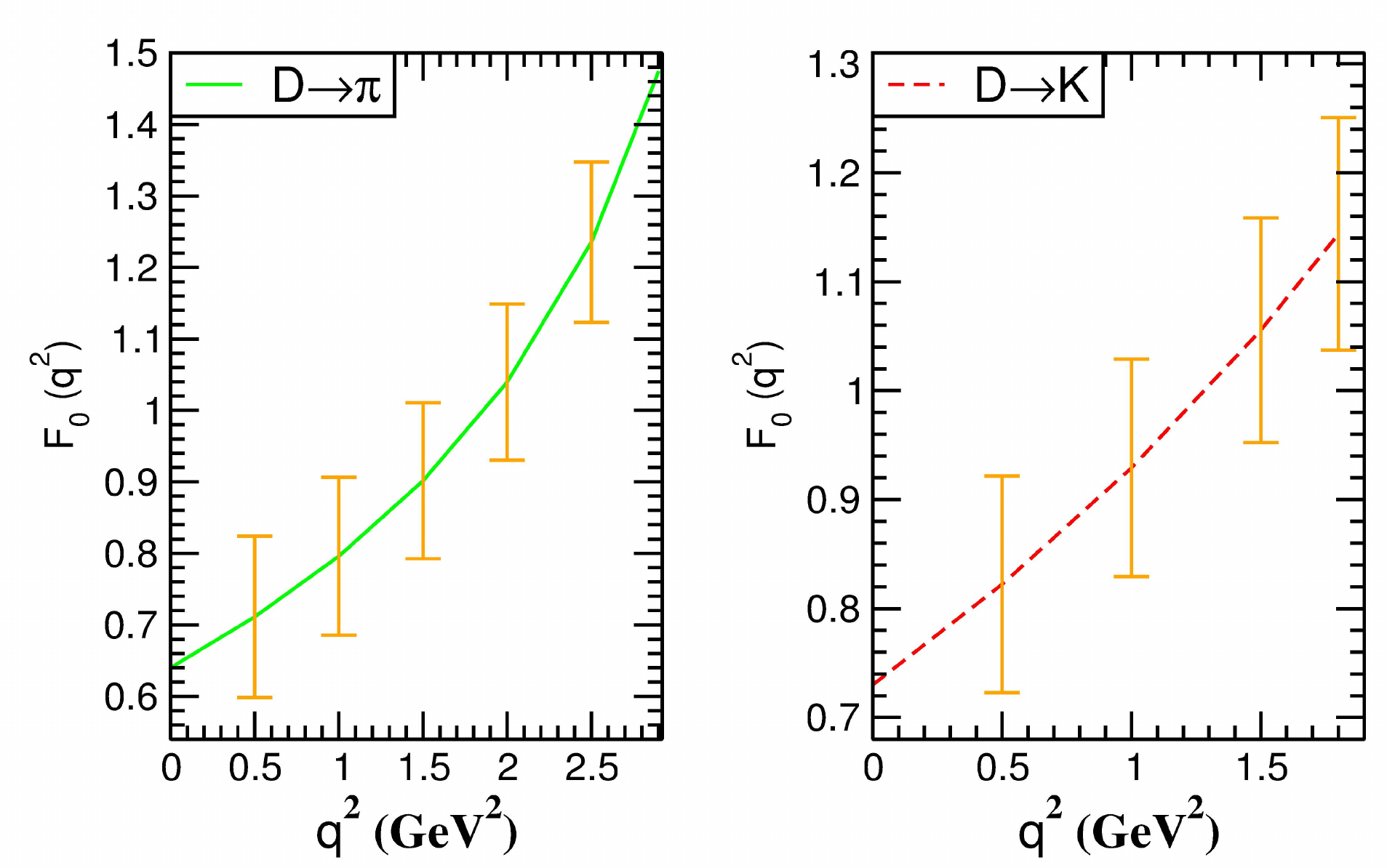}
  		\caption{The $q^2$ dependence of the scalar form factors. The plot on the left displays $F_0(q^2)$ for $D\rightarrow\pi$ transition, while that on the right is for $D\rightarrow K$ transition.}
  	        \label{plots}
        \end{center}
  \end{figure} 

 In Fig.~\ref{plots} we show the plots for our results for $D\rightarrow \pi$ and $D\rightarrow K$ where the lattice input parameters from Ref.~\cite{Bernard:2009ke} are used.
 We would like to point out that our $D\rightarrow \pi$ and $D\rightarrow K$ form factor values at $q^2=0$ are in very good agreement with that given in Refs.~\cite{PhysRevD.82.114506,PhysRevD.84.114505}
which are the most precise published calculations for $D\rightarrow\pi l\nu$ and $D\rightarrow Kl\nu$ form factors, according to Lattice Review~\cite{Aoki:2013ldr}. Further, the shape of $F_0(q^2)$ for $D\rightarrow\pi, K$ that we obtain after the z-series expansion are consistent with that of Ref.~\cite{Na:2011mc}. Moreover, the $f(0)$ values in Table~\ref{tab:BK} for $D\rightarrow\pi$ and $D\rightarrow K$ are also in agreement with CLEO results of Ref.~\cite{Besson:2009uv}. 

Regarding $D\rightarrow \eta$ and 
$D\rightarrow \eta^\prime$ form factors, since in these transitions, only the non-strange component $\eta_q$ is involved, hence one expects, $F_0(q^2)$ for $D\rightarrow \eta_q, 
\eta^\prime_q \sim D\rightarrow \pi$.
 For $D\rightarrow\eta$, CLEO has determined $F_{+}(0)V_{cd}=0.086\pm0.006\pm0.001$ using the semileptonic decay mode $D^+\rightarrow\eta e^+\nu_e$ ~\cite{Yelton:2010js}. Hence we use this value to estimate 
$F_{0}(0)$. However, for $\eta'$, while the first observation of the decay mode $D^+\rightarrow\eta'e^+\nu_e$ has been reported by CLEO in the same paper, but the form factor was not determined in this
case, and hence we approximate, the $F_{0}(0)$ for $\eta'$ to be the same as that for $\eta$. Further, since it has been shown\cite{Koponen:2012di} that the form factors and particularly their shape is insensitive to the spectator quark,  
the shape for both $\eta$ and $\eta'$, is assumed to be the same as that for the $D\rightarrow\pi$ case.

     The $D_s\rightarrow \eta, \eta^\prime$ have been estimated with some lattice studies using the ($f(0),\beta$) values from a recent exploratory paper by Bali \emph{et. al.} Ref.~\cite{Bali:2014pva}. 
However, these have larger uncertainties, since even the lowest pion mass used is still far from the physical mass. Hence, for $D_s\rightarrow\eta, \eta'$ we take the form factors to be similar to 
that of $D\rightarrow K$ and for $D_s\rightarrow K$ we take them to be similar to $D\rightarrow\pi$. Note that in all these cases the masses of the final mesons for each of the respective decay process are 
used in obtaining their z-expansion coefficients, the approximations are used only for the input parameters, $f(0)$ and $\beta$. Due to this uncertainty, we have added an additional $3\%$ theoretical 
error to these form factors.      
\begin{table}
\begin{center}
\caption{Best fit values of BK parameters for the scalar form factors}
\begin{tabular}{ |p{1.5cm} |     p{3.5cm} |  p{3.5cm}| } 
\hline\hline 
Decay                            & $ f(0)$ &  $\beta$   \\   \hline \hline
$D \rightarrow \pi$        &    $0.64 \pm 0.03 \pm 0.06$    &     $1.41 \pm 0.06 \pm 0.07$            \\   \hline
$D \rightarrow K $     &   $ 0.73 \pm  0.03 \pm 0.07$       &     $1.31 \pm 0.07 \pm 0.13$        \\   \hline
\hline 

\end{tabular}
\label{tab:BK} 
\end{center}      
\end{table}

\begin{table}
\begin{center}
\caption{ z-expansion coefficients obtained after using the BK parameters in Table~\ref{tab:BK}}
\begin{tabular}{ |p{2.0cm} |     p{2.5cm} |  p{2.5cm}| } 
\hline\hline 
Decay                            & $ a_0$ &  $a_1$  \\   \hline \hline
$D \rightarrow\pi $        &    $0.19 \pm 0.02$    &     $-0.41 \pm 0.05$            \\   \hline
$D \rightarrow K $     &   $ 0.08 \pm 0.01$       &     $-0.32 \pm 0.03$        \\   \hline
$D \rightarrow\eta  $     &   $ 0.06 \pm 0.004$       &     $-0.27 \pm 0.02$        \\   \hline
\hline 

\end{tabular}
\label{tab:zexp} 
\end{center}     
\end{table}
           
Turning now to the decay constants, for $\pi$ and $K$ mesons, the $f_{\pi, K}$  are taken from the Particle Data Group(PDG)~\cite{Agashe:2014kda}. For the $\eta$ and $\eta'$, following the method described 
in ~\cite{Feldmann:1998vh}, it is assumed that the decay constants in the quark flavour basis, follow the pattern of particle state mixing. The $\eta$ and $\eta^\prime$ are expressed as linear combinations 
of the orthogonal flavour states, 
\begin{equation}
\eta_q = \frac{1}{\sqrt{2}}(u\bar{u} + d\bar{d}), ~~~\mathrm{and}~~~\eta_s = s\bar{s}~.
\end{equation}
The physical states $\eta$ and $\eta'$ are related to these flavour states by,
\begin{eqnarray*}
	\begin{pmatrix}
		\eta\\ \eta'
	\end{pmatrix}=\begin{pmatrix}
	\cos\phi&-\sin\phi\\
	\sin\phi&\cos\phi
\end{pmatrix}\begin{pmatrix}
\eta_q\\ \eta_s
\end{pmatrix}~,
\end{eqnarray*}
where, the $\eta-\eta^\prime$ mixing angle denoted by $\phi$, represents the sum of the ideal mixing angle and the $\eta-\eta^\prime$ mixing angle ($\theta$) in the octet-singlet basis, 
$\phi = \theta+\tan^{-1}\sqrt{2}$. 
Hence the decay constants (form factors) $f_q$ and $f_s$ ($F_{0_q}$ and $F_{0_s}$) corresponding to that for $\eta_q$ and $\eta_s$ ($D\to\eta_q$ and $D\to\eta_s$) respectively, are given by:
\begin{eqnarray*}
\begin{aligned}
&f_\eta^q=f_q \cos\phi,& f_\eta^s=-f_s \sin\phi, &\qquad\qquad F_{0_\eta}^q=F_{0_q} \cos\phi,& F_{0_\eta}^s=-F_{0_s} \sin\phi,\\
&f_{\eta'}^q=f_q \sin\phi,& f_{\eta'}^s=f_s \cos\phi.&\qquad\qquad F_{0_{\eta'}}^q=F_{0_q} \sin\phi,&  F_{0_{\eta'}}^s=F_{0_s} \cos\phi.\\
\end{aligned}
\end{eqnarray*}

Various ratios of decay rates having $\eta^\prime$ in the final state with respect to that with $\eta$, for example, $\Gamma(J/\psi\to \eta^\prime \rho)/\Gamma(J/\psi\to \eta \rho)$, comparison of 
cross-sections of scattering processes for $\pi^- p\to\eta^\prime n$ with that of $\pi^- p\to\eta n$ etc., had been used for a phenomenological fit for the decay constants as well as the angle $\phi$ in 
Ref.~\cite{Feldmann:1998vh} and had been widely used. Recently, Babar with more accurate data on two photon widths of light pseudoscalar mesons, did a combined analysis ~\cite{Cao:2012nj} along with CLEO 
data to yield a mixing angle and decay constants with reduced uncertainties: $\phi=37.66\pm0.70$, $\frac{f_q}{f_{\pi}}=1.078\pm0.044$ and $\frac{f_s}{f_{\pi}}=1.246\pm0.087$, which are used in this work.

With the above inputs, the un-unitarized amplitudes for all the two-body pseudoscalar-pseudoscalar (PP) modes: SCS, CF and DCS
 may be written and are listed on the next page. For the decay modes involving $\eta$ and $\eta'$, to distinguish the case in which $\eta_q$ is the $P_2$ meson of eqn (4) from that where $\eta_s$ is the one, the notation used is:\
\begin{eqnarray*}
 C_{\eta_q}^f&=&\frac{G_F}{\sqrt{2}}V_{CKM}a_2(\mu)f_q(m_D^2-m_{P_1}^2)F_0^{DP_1}(m_{\eta}^2),\\
 C_{\eta_s}^f&=&\frac{G_F}{\sqrt{2}}V_{CKM}a_2(\mu)f_s(m_D^2-m_{P_1}^2)F_0^{DP_1}(m_{\eta}^2),\\
C_{\eta'_q}^f&=&\frac{G_F}{\sqrt{2}}V_{CKM}a_2(\mu)f_q(m_D^2-m_{P_1}^2)F_0^{DP_1}(m_{\eta'}^2),\\
C_{\eta'_s}^f&=&\frac{G_F}{\sqrt{2}}V_{CKM}a_2(\mu)f_s(m_D^2-m_{P_1}^2)F_0^{DP_1}(m_{\eta'}^2).\\
\end{eqnarray*}
\\

\begin{figure}[ht]
{\bf{SCS Decays}}
\begin{align*}
&A(D^0\rightarrow\pi^+\pi^-)=-V_{cd}V_{ud}(T + E_q)\\
&A(D^0\rightarrow\pi^0\pi^0)=\frac{V_{cd}V_{ud}}{\sqrt{2}}(-C+E_q)\\
&A(D^0\rightarrow\pi^0\eta)=\frac{V_{cd}V_{ud}}{2}(-C_{\eta_q}^{F_0}+C_{\eta_q}^f)\cos\phi-V_{cs}V_{us}C_{\eta_s}^f\frac{\sin\phi}{\sqrt{2}}-V_{cd}V_{ud} E_q\cos\phi\\
&A(D^0\rightarrow\pi^0\eta')=\frac{V_{cd}V_{ud}}{2}(-C_{\eta'_q}^{F_0}+C_{\eta'_q}^f)\sin\phi+V_{cs}V_{us}C_{\eta'_s}^f\frac{\cos\phi}{\sqrt{2}}-V_{cd}V_{ud} E_q\sin\phi\\
&A(D^0\rightarrow\eta\eta)=\frac{V_{cd}V_{ud}}{\sqrt{2}}(C_{\eta_q}^f+ 
      E_q)\cos^2\phi+V_{cs}V_{us}(-C_{\eta_s}^f\frac{\sin2\phi}{2\sqrt{2}}+
       \sqrt{2}E_s\sin^2\phi)\\
&A(D^0\rightarrow\eta\eta')=V_{cd}V_{ud}(E_q\frac{\sin2\phi}{2}+(C_{\eta_q}^f+C_{\eta'_q}^f)\frac{\sin2\phi}{4})+ 
  V_{cs}V_{us}(-C_{\eta_s}^f\frac{\sin^2\phi}{\sqrt{2}}+C_{\eta'_s}^f\frac{\cos^2\phi}{\sqrt{2}}-
    E_s\sin
      2\phi)\\
&A(D^0\rightarrow K^+K^-)=V_{cs}V_{us}(T+ 
    E_q)\\
&A(D^0\rightarrow K^0{\bar K}^0)=-(V_{cs}V_{us}E_q+ 
 V_{cd}V_{ud}E_s)\\
&A(D^+\rightarrow\pi^+\pi^0)=-\frac{V_{cd}V_{ud}}{\sqrt{2}}(T+ 
    C)\\
&A(D^+\rightarrow\pi^+\eta)=\frac{V_{cd}V_{ud}}{\sqrt{2}}(T_{\eta_q}^{F_0}+ 
    C_{\eta_q}^f+ 
    2 A_q) \cos
    \phi- 
 V_{cs}V_{us}C_{\eta_s}^f \sin\phi\\
&A(D^+\rightarrow\pi^+\eta')=\frac{V_{cd}V_{ud}}{\sqrt{2}}(T_{\eta'_q}^{F_0}+ 
    C_{\eta'_q}^f+ 
    2 A_q) \sin\phi\ + 
 V_{cs}V_{us} C_{\eta'_s}^f\cos{\phi} \\ 
&A(D^+\rightarrow K^+\bar{K}^0)=V_{cd}V_{ud} A_s + 
 V_{cs}V_{us} T \\   
&A(D_s^+\rightarrow\pi^+K^0)= -(V_{cd}V_{ud} T+V_{cs}V_{us} A_q)\\
&A(D_s^+\rightarrow\pi^0K^+)=-\frac{1}{\sqrt{2}}(V_{cd}V_{ud} 
     C-V_{cs}V_{us} A_q)\\
&A(D_s^+\rightarrow K^+\eta)=(V_{cd}V_{ud} C_{\eta_q}^f + V_{cs}V_{us} A_q)
   \frac{\cos\phi}{\sqrt{2}} - 
  V_{cs}V_{us}(T_{\eta_s}^{F_0} + 
     C_{\eta_s}^f + 
     A_s) \sin\phi\\
&A(D_s^+\rightarrow K^+\eta')=(V_{cd}V_{ud} C_{\eta'_q}^f + V_{cs}V_{us} A_q)
   \frac{\sin\phi}{\sqrt{2}} +
  V_{cs}V_{us}(T_{\eta'_s}^{F_0} + 
     C_{\eta'_s}^f + 
     A_s) \cos\phi
\end{align*}

The $D^0\rightarrow\pi\pi$ SCS decays obey the following triangular Isospin relation:
$$A(D^0\rightarrow\pi^+\pi^-)+\sqrt{2}A(D^0\rightarrow\pi^0\pi^0)=\sqrt{2}A(D^+\rightarrow\pi^+\pi^0)$$
\end{figure}

\begin{figure}[ht]
{\bf{CF Decays}}
	\begin{align*}
		&(D^0\rightarrow K^-\pi^+)= -V_{cs}V_{ud}(T+E_q)\\
		&(D^0\rightarrow\bar K^0\pi^0)= -V_{cs}V_{ud}\frac{(C-E_q)}{\sqrt{2}}\\
		&(D^0\rightarrow\bar K^0\eta)= 
		V_{cs}V_{ud}((-C_{\eta_q}^{F_0}-E_q)\frac{\cos\phi}{\sqrt{2}}+ 
		E_s\sin\phi)\\
		&(D^0\rightarrow\bar K^0\eta')= 
		V_{cs}V_{ud}((-C_{\eta'_q}^{F_0}-E_q)\frac{\sin\phi}{\sqrt{2}}-
		E_s\cos\phi)\\
		&(D^+\rightarrow\bar K^0\pi^+)= -V_{cs}V_{ud}(T+C)\\
		&(D_s^+\rightarrow\bar K^0 K^+)=- V_{cs}V_{ud}(C+A_s)\\
		&(D_s^+\rightarrow\pi^+\eta)= \frac{V_{cs}V_{ud}}{\sqrt{2}}(T_{\eta_s}^{F_0}\sin\phi-A_q\cos\phi)\\
		&(D_s^+\rightarrow\pi^+\eta')=\frac{V_{cs}V_{ud}}{\sqrt{2}}(-T_{\eta'_s}^{F_0}\cos\phi-A_q\sin\phi)
	\end{align*}

The $D^0\rightarrow K\pi$ CF decays obey the following triangular Isospin relation:
$$A(D^0\rightarrow K^-\pi^+)+\sqrt{2}A(D^0\rightarrow\bar{K^0}\pi^0)=A(D^+\rightarrow\bar{K^0}\pi^+)$$

{\bf{DCS Decays}}
	\begin{align*}
		&(D^0\rightarrow K^+\pi^-)= V_{cd}V_{us}(T+E_q)\\
		&(D^0\rightarrow K^0\pi^0)= V_{cd}V_{us}\frac{(C-E_q)}{\sqrt{2}}\\
		&(D^0\rightarrow K^0\eta)= 
		V_{cd}V_{us}((C_{\eta_q}^{F_0}+E_q)\frac{\cos\phi}{\sqrt{2}}- 
		E_s\sin\phi)\\
		&(D^0\rightarrow K^0\eta')= 
		V_{cd}V_{us}((C_{\eta'_q}^{F_0}+E_q)\frac{\sin\phi}{\sqrt{2}}+
		E_s\cos\phi)\\
		&(D^+\rightarrow K^0\pi^+)= V_{cd}V_{us}(C+A_q)\\
		&(D^+\rightarrow K^+\pi^0)= V_{cd}V_{us}\frac{T-A_q}{\sqrt2}\\
		&(D^+\rightarrow K^+\eta)= -V_{cd}V_{us}((T_{\eta_q}^{F_0}+A_q)\frac{\cos\phi}{\sqrt{2}}-A_s\sin\phi)\\
		&(D^+\rightarrow K^+\eta')= -V_{cd}V_{us}((T_{\eta'_q}^{F_0}+A_q)\frac{\sin\phi}{\sqrt{2}}+A_s\cos\phi)\\
		&(D_s^+\rightarrow K^+ K^0)= V_{cd}V_{us}(T+C)
	\end{align*}

The $D^0\rightarrow K\pi$ DCS decays obey the following quadrilateral Isospin relation:
$$A(D^0\rightarrow K^+\pi^-)+\sqrt{2}A(D^0\rightarrow K^0\pi^0)=A(D^+\rightarrow K^0\pi^+)+\sqrt{2}A(D^+\rightarrow K^+\pi^0)$$
\end{figure}

Similarly, to distinguish the cases where $\eta_q$ or $\eta_s$ is the $P_1$ meson, which incidentally appears in both Tree($T$) and Colour Suppressed($C$) amplitudes (unlike for the case discussed above), 
we use the notation: 
\begin{eqnarray*}
 C_{\eta_q}^{F_0}(T_{\eta_q}^{F_0})&=&\frac{G_F}{\sqrt{2}}V_{CKM}a_2(\mu)(a_1(\mu))f_{P_2}(m_D^2-m_{\eta}^2)F_0^{D\eta_q}(m_{P_2}^2),\\
 C_{\eta_s}^{F_0}(T_{\eta_s}^{F_0})&=&\frac{G_F}{\sqrt{2}}V_{CKM}a_2(\mu)(a_1(\mu))f_{P_2}(m_D^2-m_{\eta}^2)F_0^{D\eta_s}(m_{P_2}^2),\\
 C_{\eta'_q}^{F_0}(T_{\eta'_q}^{F_0})&=&\frac{G_F}{\sqrt{2}}V_{CKM}a_2(\mu)(a_1(\mu))f_{P_2}(m_D^2-m_{\eta'}^2)F_0^{D\eta'_q}(m_{P_2}^2),\\
 C_{\eta'_s}^{F_0}(T_{\eta'_s}^{F_0})&=&\frac{G_F}{\sqrt{2}}V_{CKM}a_2(\mu)(a_1(\mu))f_{P_2}(m_D^2-m_{\eta'}^2)F_0^{D\eta'_s}(m_{P_2}^2).\\
\end{eqnarray*}

\section{Final State Interactions}\label{Final State Interactions}
Final state interaction effects are incorporated using unitarity relations, where the contribution to any channel is a result of sum over all possible hadronic intermediate states. Hence for all the 
$n$ $D\rightarrow PP$ decays, the FSI corrected amplitudes or the 'unitarized' amplitudes, $\cal{A}_{\mathrm{i}}^{\mathrm{U}}$ with $i=1,...n$ are given by~\footnote{For part of the formalism used in this section,
 we closely follow Refs.~\cite{Chua:2008kp,Chua:2005dt}} 
\begin{equation}
\cal{A}_\mathrm{i}^\mathrm{U} = \displaystyle \sum_{\mathrm{k=1}}^\mathrm{N} \cal{S}_{\mathrm{ik}}^{\mathrm{1/2}}\cal{A}_\mathrm{k}~,
\label{FullScatt}
\end{equation}
where, $\cal{S}$ is the strong interaction matrix and $k=1,...n,n+1,....N$, stands for all possible states that can rescatter into the $PP$ states.  
In the heavy quark limit the hard rescattering dominates, in which case, the sum can be interpreted to be over all intermediate states of partons and the number of these states will hence be very large. 
Parton hadron duality will then permit this estimation. These corrections are incorporated into the hard scattering contributions in the QCD factorization approach of Ref.~\cite{Beneke:2003zv} 
for the case of B meson decays. For the case of charmed meson decays, since $m_c$ is not large enough, we include these NF corrections in the modified Wilson coefficients, in terms of 
parameters $\chi_1$ and $\chi_2$. However, some residual long distance FSI's may be left which are particularly important for charmed meson decays, due to the nearby resonances. 
This residual rescattering is considered in the limited set of $D\rightarrow PP$ decays, to which the duality cannot be applied and therefore these effects may not be incorporated in the NF 
corrections.

The $\cal{S}$ matrix in Eq.~(\ref{FullScatt}) can be written in terms of a residual matrix ($\cal{S}_{\mathrm{res}}$) for the rescattering among the $D\rightarrow PP$ states alone and the scattering 
matrix which accounts for the hard rescattering from all possible hadronic states  into these channels, resulting in the factorization amplitudes as, 
\begin{eqnarray}
\cal{S}_{\mathrm{ik}} &=& \displaystyle\sum_{\mathrm{j=1}}^{\mathrm{n}} (\cal{S}_\mathrm{1})_{\mathrm{ij}}(\cal{S}_\mathrm{2})_{\mathrm{jk}}, ~\mathrm{where},~ \cal{S}_\mathrm{1} =\cal{S}_{\mathrm{res}}~  
\mathrm{and}\nonumber\\ \cal{S}_\mathrm{2}&=&\cal{S}_\mathrm{1}^{\mathrm{-1}}\cal{S}, \mathrm{with}~
\cal{A}_\mathrm{j}^{\mathrm{fac}} = \displaystyle\sum_{\mathrm{k=1}}^\mathrm{N} (\cal{S}_\mathrm{2}^{\mathrm{1/2}})_{\mathrm{jk}}\cal{A}_\mathrm{k},~ \mathrm{resulting~in}\nonumber\\ 
\cal{A}_\mathrm{i}^\mathrm{U} &=& \displaystyle \sum_{\mathrm{k=1}}^{\mathrm{n}} (\cal{S}_\mathrm{res}^{\mathrm{1/2}})_{\mathrm{ij}}\cal{A}_\mathrm{j}^{\mathrm{fac}}.~
\label{scat}
\end{eqnarray}
Any $\cal{S}$ matrix can be written in terms of a real, symmetric K matrix as, $\cal{S}=\mathrm{(1-iK)^{-1}(1+iK)}$. Hence, the unitarized amplitudes in Eq.~(\ref{scat}), may be written as,
\begin{equation}
 \cal{A}_\mathrm{i}^\mathrm{U} = \displaystyle \sum_{\mathrm{k=1}}^{\mathrm{n}} (\mathrm{(1-iK)^{-1}})_{\mathrm{ij}}\cal{A}_\mathrm{j}^{\mathrm{fac}}~.
 \label{summation}
\end{equation}
The K matrix parametrization has the advantage that the resonances coupling two body channels are represented by poles in the K matrix. The summation in Eq.~(\ref{summation}), corresponds to summing the 
geometric series, where the final state hadrons are produced from scattering via resonance at different orders, starting from zero, i.e., directly from the decaying meson without the resonance contribution, 
resonant rescattering occurring once, twice and so on. While such coupled FSI's has been considered in the past in many papers~\cite{Sorensen:1981vu,Kamal:1988ub,Zenczykowski:1996xd,Kamal:1987nm,Buccella:2013tya} most 
of the papers on charm decays further assume SU(3) to relate the parameters of the coupling matrix.
Since SU(3) is broken, we prefer to use the measured decay rates of the resonances to various channels to fix the K matrix parameters as far as possible and the ones not measured are left as 
parameters to be determined by fits of all the theoretical branching ratios to the observed values. 
  
For each of the SCS, CF as well as DCS modes, states with the same isospin are coupled together. In general the $K$ matrix coupling 
three channels will have the form:
\begin{eqnarray*}
	K(s)=\frac{1}{(m_{Res}^2-s)}\begin{bmatrix}
		k_1\Gamma_{11}&\sqrt{k_1k_2}\Gamma_{12}&\sqrt{k_1k_3}\Gamma_{13}\\ \sqrt{k_2k_1}\Gamma_{21}&k_2\Gamma_{22}&\sqrt{k_2k_3}\Gamma_{23}\\ 
\sqrt{k_3k_1}\Gamma_{31}&\sqrt{k_3k_2}\Gamma_{32}&k_3\Gamma_{33},
	\end{bmatrix} 
\end{eqnarray*}
where, $m_{Res}$ denotes the mass of the resonance through which the different channels are coupled and $k_1$, $k_2$ and $k_3$ are the cm momenta of the 3 decay modes. There are six independent parameters 
$\Gamma_{ij}$. To reduce the independent parameters to a manageable number, we impose the requirement that the diagonal cofactors of $K(s)$ vanish (or equivalently, det$K(s)=0$). This leads us to three 
conditions,
\begin{equation}
	\Gamma_{12}^2=\Gamma_{11}\Gamma_{22}~,\;\;\;\;\; \Gamma_{13}^2=\Gamma_{11}\Gamma_{33} \;\;\; \mathrm{and}  \;\;\; \Gamma_{23}^2=\Gamma_{22}\Gamma_{33}.~
	\label{K-const}
\end{equation} 
The $\Gamma_{ii}'s$ are related to the the partial decay width of the resonance to the $i^{th}$ channel. 

To illustrate this, we consider first the case of isospin zero states of SCS decay modes of $D^0$ meson. The  isospin zero  combination of the $\pi^+\pi^-$ and $\pi^0\pi^0$, $K^+K^-$ and $K^0\bar{K}^0$, and 
the $\eta\eta$ modes, are coupled via Eq.~(\ref{summation}) with the $f_0(1710)$ pole in the K matrix. Hence, for this specific case of coupling of the isospin zero states, in the K matrix, $m_{Res}=1.720$ 
GeV,  $k_1=\frac{1}{2}({m_{D^0}}^2-4m_{\pi^0}^2)$, $k_2=\frac{1}{2}({m_{D^0}}^2-4m_{K^0}^2)$, $k_3=\frac{1}{2}({m_{D^0}}^2-4m_{\eta}^2)$~ 
and we have 
\begin{equation*}
 \Gamma(f_0\rightarrow\pi\pi)=\frac{\Gamma_{11}k_1}{m_{Res}},\;\;\;\;\;
 \Gamma(f_0\rightarrow K\bar K)=\frac{\Gamma_{22}k_2}{m_{Res}},\;\;\;\;\;\mathrm{and}\;\;\;\;\;
 \Gamma(f_0\rightarrow\eta\eta)=\frac{\Gamma_{33}k_3}{m_{Res}}.
\end{equation*}
Experimentally only the two ratios of the decay rates,  $\Gamma(f_0\rightarrow K\bar K)/\Gamma(f_0\rightarrow\pi\pi)$ and $\Gamma(f_0\rightarrow K\bar K)/\Gamma(f_0\rightarrow\eta\eta)$ have been determined.
 Hence we keep $g_{pe}\equiv\Gamma(f_0\rightarrow K\bar K)$  as a parameter, to be determined from fits of our theoretical estimates to the observed branching ratios.
 
Similarly, for the I=1 case, we take the $a_0(1450)$ resonance with $m_{Res}=1.474$ GeV and $\Gamma_{Res}=0.265$ GeV, to be responsible for the rescattering among the  channels $K\bar{K}$, $\pi\eta$ and 
$\pi\eta'$, to which this resonance decays. Here again, the decay rate $\Gamma(a_0\rightarrow\pi\eta)$ is not yet accurately measured and is treated as a parameter $(h_{pe})$ that may be predicted from the 
fits of all the branching ratios of the $D\rightarrow PP$ modes to experimental data. Note that the $K\bar{K}$, $\pi\eta$ and $\pi\eta'$ states appear as final states not only of SCS $D^0$ and $D^+$ decays, 
but also in the CF decays of the $D_s^+$ decays. For all these three sets of decays, the same K matrix (apart from tiny modifications in the cm momenta and the mass-squared of the decaying meson) will 
suffice, and more importantly with the same one unknown parameter, while many additional observables (all the branching ratios of these $D^0$, $D^+$ and $D_s^+$) will get added to the $\chi^2$ fit. If this 
one same parameter, along with the other unknowns in our analysis can simultaneously explain all the observed data, it would indicate that our naive technique of incorporating the FSI effects is satisfactory.  

We also couple the isospin $1/2$ states of the $K\pi$, $K\eta$ and $K\eta^\prime$ channels, that are the final states in the SCS decays of $D_s^+$, CF decays of $D^0$ and DCS decays of $D^0$ and $D^+$
 mesons. Here we use the $K_0^*(1950)$ resonance with $m_{Res} = 1.945$ GeV. Only the branching ratio, $\Gamma(K_0^*\rightarrow K\pi)/\Gamma_{total}$ has been measured. We take the other two decay rates, 
$\Gamma(K_0^*\rightarrow K\eta)$ and  $\Gamma(K_0^*\rightarrow K\eta^\prime)$ as parameters ($j_{pe_1}$ and $j_{pe_2}$) that can be determined by the overall fits of all the branching ratios to data. 
 
\section{Numerical Analysis and Results}\label{Numerical Analysis}
To estimate all the possible sets of coupled channels, the isospin decomposition of all the SCS, CF and DCS modes are listed on the next page.
Here  $A^{(U)}$  denote the bare or un-unitarized (unitarized or FSI corrected) amplitudes respectively, for each of the decay modes, while $A^{mode(U)}_i $ denotes the 
corresponding ununitarized (unitarized) isospin, $I=i$ amplitudes for those modes. With absence of resonances in particular isopin components with the right quantum numbers in the vicinity of the charmed 
meson masses, some of the isospin components of many modes remain un-unitarized.
\\

\begin{figure}[ht]
{\bf{SCS Decays}}
\begin{align*}
&A^{(U)}(D^0\rightarrow\pi^+\pi^-)\equiv\sqrt{2}\mathcal{A}_2^{\pi\pi}+\sqrt{2}\mathcal{A}_0^{\pi\pi (U)}\\
&A^{(U)}(D^0\rightarrow\pi^0\pi^0)\equiv2\mathcal{A}_2^{\pi\pi}-\mathcal{A}_0^{\pi\pi (U)}\\
&A^{(U)}(D^0\rightarrow\pi^0\eta)\equiv\sqrt{3}\mathcal{A}_1^{\pi\eta (U)}\\
&A^{(U)}(D^0\rightarrow\pi^0\eta)\equiv\sqrt{3}\mathcal{A}_1^{\pi\eta' (U)}\\
&A^{(U)}(D^0\rightarrow\eta\eta)\equiv\sqrt{3}\mathcal{A}_0^{{\eta\eta} (U)}\\
&A^{(U)}(D^0\rightarrow\eta\eta')\equiv\sqrt{3}\mathcal{A}_0^{\eta\eta'}\\
&A^{(U)}(D^0\rightarrow K^+K^-)\equiv\sqrt{\frac{3}{2}}\left(\mathcal{A}_{1}^{KK (U)}+\mathcal{A}_0^{KK (U)}\right)\\
&A^{(U)}(D^0\rightarrow K^0\bar{K}^0)\equiv\sqrt{\frac{3}{2}}\left(\mathcal{A}_{1}^{KK (U)}-\mathcal{A}_0^{KK (U)}\right)\\
&A^{(U)}(D^+\rightarrow\pi^+\pi^0)\equiv3\mathcal{A}_2^{\pi\pi}\\
&A^{(U)}(D^+\rightarrow K^+\bar{K}^0)\equiv\mathcal{A}_{1}^{K^+K (U)}\\
&A^{(U)}(D^+\rightarrow\pi^+\eta)\equiv\mathcal{A}_{1}^{\pi^+\eta (U)}\\
&A^{(U)}(D^+\rightarrow\pi^+\eta')\equiv\mathcal{A}_{1}^{\pi^+\eta' (U)}\\
&A^{(U)}(D_s^+\rightarrow\pi^+K^0)\equiv\frac{1}{\sqrt{3}}\mathcal{A}_{\frac{3}{2}}^{\pi K}
                         +\sqrt{\frac{2}{3}}\mathcal{A}_{\frac{1}{2}}^{\pi K(U)}\\
&A^{(U)}(D_s^+\rightarrow\pi^0K^+)\equiv\sqrt{\frac{2}{3}}\mathcal{A}_{\frac{3}{2}}^{\pi K}
                         -\frac{1}{\sqrt{3}}\mathcal{A}_{\frac{1}{2}}^{\pi K(U)}\\
&A^{(U)}(D_s^+\rightarrow K^+\eta)\equiv\mathcal{A}_{\frac{1}{2}}^{K^+\eta(U)}\\
&A^{(U)}(D_s^+\rightarrow K^+\eta')\equiv\mathcal{A}_{\frac{1}{2}}^{K^+\eta'(U)}\\
\end{align*}

{\bf{CF Decays}}
        \begin{align*}
		&A^{(U)}(D^0\rightarrow K^-\pi^+)= \frac{1}{3}\mathcal{A}_{\frac{3}{2}}^{\bar K\pi} + 
		\frac{2}{3}\mathcal{A}_{\frac{1}{2}}^{\bar K\pi (U)}\\
		&A^{(U)}(D^0\rightarrow\bar K^0\pi^0)= \frac{\sqrt{2}}{3}(\mathcal{A}_{\frac{3}{2}}^{\bar K\pi}-
		\mathcal{A}_{\frac{1}{2}}^{\bar K\pi (U)})\\
		&A^{(U)}(D^0\rightarrow\bar K^0\eta)= 
		\sqrt{\frac{2}{3}}\mathcal{A}_{\frac{1}{2}}^{\bar K\eta (U)}\\
		&A^{(U)}(D^0\rightarrow\bar K^0\eta')= 
		\sqrt{\frac{2}{3}}\mathcal{A}_{\frac{1}{2}}^{\bar K\eta' (U)}\\
        \end{align*}
\end{figure}

\begin{figure}[ht]
        \begin{align*}
		&A^{(U)}(D^+\rightarrow\bar K^0\pi^+)= {A}_{\frac{3}{2}}^{\bar K\pi^+}\\
		&A^{(U)}(D_s^+\rightarrow\bar K^0 K^+)= \mathcal{A}_{1}^{K\bar K (U)}\\
		&A^{(U)}(D_s^+\rightarrow\pi^+\eta)= \mathcal{A}_{1}^{\pi^+\eta (U)}\\
		&A^{(U)}(D_s^+\rightarrow\pi^+\eta')=\mathcal{A}_{1}^{\pi^+\eta' (U)}
	\end{align*}

{\bf{DCS Decays}}
        \begin{align*}
		&A^{(U)}(D^0\rightarrow K^+\pi^-)= \frac{\sqrt{2}}{3}\mathcal{A}_{\frac{3}{2}}^{K\pi}- 
		\frac{\sqrt{2}}{\sqrt{3}}\mathcal{A}_{\frac{1}{2}}^{K\pi (U)}\\
		&A^{(U)}(D^0\rightarrow K^0\pi^0)= \frac{2}{3}\mathcal{A}_{\frac{3}{2}}^{K\pi}+
		\frac{1}{\sqrt{3}}\mathcal{A}_{\frac{1}{2}}^{K\pi (U)}\\
		&A^{(U)}(D^0\rightarrow K^0\eta)= \mathcal{A}_{\frac{1}{2}}^{K\eta (U)}\\
		&A^{(U)}(D^0\rightarrow K^0\eta')= \mathcal{A}_{\frac{1}{2}}^{K\eta' (U)}\\
		&A^{(U)}(D^+\rightarrow K^0\pi^+)= \frac{\sqrt{2}}{3}\mathcal{A}_{\frac{3}{2}}^{K\pi^+}+ 
		\frac{\sqrt{2}}{\sqrt{3}}\mathcal{A}_{\frac{1}{2}}^{K\pi^+ (U)}\\
		&A^{(U)}(D^+\rightarrow K^+\pi^0)= \frac{2}{3}\mathcal{A}_{\frac{3}{2}}^{K\pi^+}-
		\frac{1}{\sqrt{3}}\mathcal{A}_{\frac{1}{2}}^{K\pi^+ (U)}\\
		&A^{(U)}(D^+\rightarrow K^+\eta)= {A}_{\frac{1}{2}}^{K^+\eta (U)}\\
		&A^{(U)}(D^+\rightarrow K^+\eta')= {A}_{\frac{1}{2}}^{K^+\eta' (U)}\\
		&A^{(U)}(D_s^+\rightarrow K^+ K^0)= \frac{1}{\sqrt{2}}(\mathcal{A}_{1}^{KK}+\mathcal{A}_{0}^{KK})
	\end{align*} 
\end{figure}

 With all the unitarized isospin amplitudes, we construct the corresponding unitarized decay amplitudes for all the decay modes. The decay rates for all the $D\rightarrow P_1P_2$ are then calculated as,
 \begin{eqnarray}
 \Gamma(D\rightarrow P_1P_2)=\frac{p_c}{8\pi m_D^2}|A(D\rightarrow P_1P_2)|^2.
 \end{eqnarray}
 Here $p_c$ is the centre of mass momentum of the mesons in the final state given by 
\begin{equation*}
 p_c=\frac{\sqrt{(m_D^2-(m_{P_1}+m_{P_2})^2)(m_D^2-(m_{P_1}-m_{P_2})^2)}}{2m_D}. 
\end{equation*}
The theoretical branching ratios for each of the decay modes of the $D^0$, $D^+$ or the $D_s^+$ mesons are then obtained by dividing the corresponding decay rates by the total decay widths of these mesons. We then perform 
a $\chi^2$ fit of these theoretical branching ratios with the experimentally measured branching fractions, estimating all the unknown parameters from the best fit to data.

The unknown parameters in our study are: the four parameters representing the NF corrections, $\chi_1$, $\chi_2$ and their respective phases $\phi_1$, $\phi_2$, four parameters: $\chi_{q,s}^{E}$
 and $\chi_{q,s}^{A}$ depicting the strength of the W-exchange and W-annihilation amplitudes with distinct strengths for $q\bar{q}$ and $s\bar{s}$ pair production, one unknown in each of the isospin zero and 
isospin one K matrices coupling modes from decays of $D^0$ and $D^+$ mesons, two parameters in the isospin half K matrix coupling various decay modes of $D_s^+$, one parameter $\Lambda$, representing the 
momentum of the soft degrees of freedom in the charmed mesons, that is used to define the scale for each of the individual decay modes, making a total of 13 unknown parameters. On the other hand out of all 
the 33 decay modes considered, 28 have been measured, resulting in sufficient observables to deterimine all the unknown parameters and give predictions for five of the branching fractions of DCS 
modes that are not yet measured.

Apart from the experimental errors in the observed branching ratios, the calculated errors in our theoretical branching ratio estimates arise from the errors in the form factors, the 
$\eta$ and $\eta^\prime$  decay constants, the $\eta-\eta^\prime$ mixing angle and the errors in the measured decay widths of the various resonances into the different channels that are included in our 
$\chi^2$ fits. Errors due to the other theoretical inputs, like meson masses, decay constants of pion, Kaon, charmed mesons, CKM mixing elements (involving the first two generations) are negligibly small.

As discussed in Sec.~\ref{FF}, the z-series expansion has been obtained for the form factors, keeping the first two terms of this series. The coefficients $a_0$ and $a_1$ of these two terms are functions 
of normalization and shape parameters $f(0)$ and $\beta$, obtained from lattice results. 
The errors in these lattice parameters are used to obtain the errors in the expansion coefficient functions and then propagated to get the errors in the form factors.
We find that the errors in the form factors vary from $\approx$5\%  to $\approx$25\%.

In the tables~\ref{table:BRvalSCS}, ~\ref{table:BRvalCF} and ~\ref{table:BRvalDCS}, we list the values of the Branching Ratios of the all the SCS, CF and DCS $D\rightarrow PP$ modes obtained from our 
analysis, after incorporating the FSI effects (shown in the $2^{nd}$ column), as well as in the absence of the FSI (column 3), absence of annihilation (column 4) along with the corresponding 
observed experimental branching ratios (column 5) given in PDG~\cite{Agashe:2014kda}. We also predict the B.R.'s for a few DCS modes that have not been experimentally measured yet and are given in the second 
column of Table~\ref{table:BRvalDCS}

After incorporating all the errors, the $\chi^2$ minimization results in the following best fit values of all the parameters:
\begin{table}[ht]
	\caption{Parameter best fit values} 
	\centering 
	\begin{tabular}{|p{1.0cm} |     p{2.3cm}|     p{1.0cm}|     p{2.3cm}|     p{1.0cm}|     p{2.3cm}| } 
		\hline\hline 
		Name & Values & Name & Values&Name&Values  \\ [0.5ex] 
		\hline\hline 
		$\Lambda$    &0.625645&$j_{{pe}_1}$ &0.0000239368 &$\chi_{q}^A$&132.685 \\ \hline 
		$\chi_1$     &-2.68215& $j_{{pe}_2}$&0.096456  &$\chi_{s}^A$&193.447  \\     \hline
		$\chi_2$     &2.23605 & $\chi_{q}^E$&-334.805 &$\phi_1$&0.302258 \\      \hline 
		$g_{pe}$     &0.0471262&$\chi_{s}^E$ &-81.3363&$\phi_2$&2.87681 \\   \hline
		$h_{pe}$     &0.118834  &             &        &        &         \\    \hline                        
        \end{tabular}
	\label{table:paraval} 
\end{table}

\begin{table}[ht]
	\caption{$D\rightarrow PP$ SCS B.R.'s, Columns 2 and 3 show our results  with annihilation included, for the with and without FSI cases respectively, while column 4 which displays results without annihilation, includes FSI.} 
	\centering 
	\begin{tabular}{ |p{2.5cm} |     p{3.5cm} |     p{3.5cm} |     p{3.5cm}|  p{3.5cm} |  }  
		\hline\hline 
		Modes & With FSI &Without FSI&Without Ann &Experimental Value \\ [0.5ex] 
		\hline\hline 
		$D^0\rightarrow\pi^+\pi^-$   &$(1.44\pm0.027)\times10^{-3}$ &$(4.35\pm1.67)\times10^{-3}$&$(4.02\pm1.75\times)\times10^{-3}$&$(1.402\pm0.026)\times10^{-3}$ \\ \hline
		$D^0\rightarrow\pi^0\pi^0$   &$(1.14\pm0.56)\times10^{-3}$&$(3.66\pm1.43)\times10^{-3}$&$(2.04\pm0.79)\times10^{-3}$&$(8.209\pm0.35)\times10^{-4}$ \\     \hline
		$D^0\rightarrow K^+K^-$       &$(4.06\pm0.77)\times10^{-3}$&$(4.27\pm2.34)\times10^{-3}$&$(6.78\pm3.08)\times10^{-3}$&$(3.96\pm0.08)\times10^{-3}$ \\   \hline
		$D^0\rightarrow K^0\bar{K^0}$ &$(3.42\pm0.52)\times10^{-4}$&$(5.61\pm0.00)\times10^{-4}$&$(2.80\pm0.84)\times10^{-4}$&$(3.4\pm0.8)\times10^{-4}$ \\     \hline
		$D^0\rightarrow\pi^0\eta$    &$(1.47\pm0.90)\times10^{-3}$&$(6.47\pm2.98\times10^{-3}$&$(3.25\pm1.51\times10^{-3}$&$(6.8\pm0.7)\times10^{-4}$ \\     \hline
		$D^0\rightarrow\pi^0\eta'$   &$(2.17\pm0.65)\times10^{-3}$&$(3.81\pm1.43)\times10^{-3}$&$(1.85\pm0.79)\times10^{-3}$&$(9.0\pm1.4)\times10^{-4}$ \\     \hline
		$D^0\rightarrow\eta\eta$     &$(1.27\pm0.27)\times10^{-3}$&$(1.32\pm0.41)\times10^{-3}$&$(1.34\pm0.29)\times10^{-3}$&$(1.67\pm0.20)\times10^{-3}$ \\   \hline
		$D^0\rightarrow\eta\eta'$    &$(9.53\pm1.83)\times10^{-4}$&$(1.04\pm0.27)\times10^{-3}$&$(5.38\pm1.63)\times10^{-4}$&$(1.05\pm0.26)\times10^{-3}$ \\   \hline
		$D^+\rightarrow\pi^+\pi^0$   &$(8.89\pm4.51)\times10^{-4}$&$(8.70\pm6.70)\times10^{-4}$&$(9.73\pm3.94)\times10^{-4}$&$(1.19\pm0.06)\times10^{-3}$ \\   \hline
		$D^+\rightarrow K^+\bar{K^0}$ &$(3.75\pm0.63)\times10^{-3}$&$(1.02\pm0.37)\times10^{-2}$&$(1.99\pm0.56)\times10^{-2}$&$(5.66\pm0.32)\times10^{-3}$ \\   \hline
		$D^+\rightarrow\pi^+\eta$    &$(4.72\pm0.21)\times10^{-3}$&$(2.34\pm1.26)\times10^{-2}$&$(1.66\pm0.77)\times10^{-2}$&$(3.53\pm0.21)\times10^{-3}$ \\   \hline
		$D^+\rightarrow\pi^+\eta'$   &$(6.76\pm2.19)\times10^{-3}$&$(3.00\pm0.76)\times10^{-2}$&$(9.78\pm3.35)\times10^{-3}$&$(4.67\pm0.29)\times10^{-3}$ \\   \hline
		$D_s^+\rightarrow\pi^+ K^0$   &$(1.96\pm0.90)\times10^{-3}$&$(1.46\pm1.10)\times10^{-3}$&$(1.32\pm1.01)\times10^{-3}$&$(2.42\pm0.12)\times10^{-3}$ \\   \hline
		$D_s^+\rightarrow\pi^0 K^+$   &$(8.17\pm4.64)\times10^{-4}$&$(1.74\pm1.00)\times10^{-4}$&$(1.01\pm0.54)\times10^{-3}$&$(6.3\pm2.1)\times10^{-4}$ \\   \hline
		$D_s^+\rightarrow K^+\eta$   &$(1.50\pm0.75)\times10^{-3}$&$(6.40\pm4.52)\times10^{-3}$&$(2.23\pm1.82)\times10^{-3}$&$(1.76\pm0.35)\times10^{-3}$ \\   \hline
		$D_s^+\rightarrow K^+\eta'$   &$(7.07\pm0.49)\times10^{-4}$&$(2.09\pm0.87)\times10^{-3}$&$(0.57\pm0.47)\times10^{-4}$&$(1.8\pm0.6)\times10^{-3}$ \\   \hline
		\hline 
	\end{tabular}
	\label{table:BRvalSCS} 
\end{table}

\begin{table}[ht]
	\caption{$D\rightarrow PP$ CF B.R.'s, inclusion of Annihilation/FSI in our branching ratio estimates shown in various columns is the same as specified for Table~\ref{table:BRvalSCS}. } 
	\centering 
	\begin{tabular}{ |p{2.5cm} |     p{3.5cm} |     p{3.5cm} |    p{3.5cm}|  p{3.5cm} |  }  
		\hline\hline 
		Modes & With FSI &Without FSI&Without Ann & Experimental Value \\ [0.5ex] 
		\hline\hline 
		$D^0\rightarrow K^-\pi^+$      &$(3.70\pm1.33)\times10^{-2}$&$(8.83\pm2.47)\times10^{-2}$&$(5.63\pm1.81)\times10^{-2}$&$(3.88\pm0.05)\times10^{-2}$ \\ \hline
		$D^0\rightarrow\bar{K^0}\pi^0$ &$(1.88\pm0.99)\times10^{-2}$ &$(1.29\pm0.44)\times10^{-1}$ &$(3.30\pm1.47)\times10^{-2}$&$(2.38\pm0.08)\times10^{-2}$ \\     \hline
		$D^0\rightarrow\bar{K^0}\eta$  &$(1.59\pm0.48)\times10^{-2}$&$(0.97\pm0.33)\times10^{-2}$ &$(1.09\pm0.34)\times10^{-2}$&$(0.958\pm0.06)\times10^{-2}$ \\   \hline
		$D^0\rightarrow\bar{K^0}\eta'$ &$(2.29\pm0.43)\times10^{-2}$&$(2.06\pm0.30)\times10^{-2}$&$(2.45\pm0.47)\times10^{-2}$&$(1.88\pm0.1)\times10^{-2}$ \\     \hline
		$D^+\rightarrow\bar{K^0}\pi^+$ &$(3.42\pm1.78)\times10^{-2}$ &$(1.35\pm1.12)\times10^{-1}$ &$(5.25\pm3.34)\times10^{-2}$&$(2.94\pm0.14)\times10^{-2}$ \\     \hline
		$D_s^+\rightarrow\bar{K^0}K^+$   &$(5.65\pm1.29)\times10^{-2}$&$(1.70\pm0.79)\times10^{-1}$ &$(1.35\pm0.53)\times10^{-1}$&$(2.95\pm0.14)\times10^{-2}$ \\   \hline
		$D_s^+\rightarrow\pi^+\eta$      &$(2.26\pm0.82)\times10^{-2}$ &$(0.78\pm0.56)\times10^{-2}$ &$(2.14\pm0.90)\times10^{-2}$&$(1.69\pm0.10)\times10^{-2}$ \\   \hline
		$D_s^+\rightarrow\pi^+\eta'$     &$(2.64\pm0.78)\times10^{-2}$&$(3.73\pm1.52)\times10^{-2}$&$(2.52\pm0.85)\times10^{-2}$&$(3.94\pm0.25)\times10^{-2}$ \\   \hline
		\hline 
	\end{tabular}
	\label{table:BRvalCF} 
\end{table}

\begin{table}[ht]
	\caption{$D\rightarrow PP$ DCS B.R.'s, inclusion of Annihilation/FSI in our branching ratio estimates shown in various columns is the same as specified for Table~\ref{table:BRvalSCS}.} 
	\centering 
	\begin{tabular}{ |p{2.5cm} |     p{3.5cm} |     p{3.5cm}|   p{3.5cm}|  p{3.5cm} |  }  
		\hline\hline 
		Modes & With FSI &Without FSI&Without Ann & Experimental Value \\ [0.5ex] 
		\hline\hline 
		$D^0\rightarrow K^+\pi^-$ &$(1.77\pm0.88)\times10^{-4}$ &$(3.71\pm1.33)\times10^{-4}$&$(2.48\pm1.07)\times10^{-4}$&$(1.38\pm0.028)\times10^{-4}$ \\ \hline
		$D^0\rightarrow K^0\pi^0$ &$(2.11\pm0.26)\times10^{-4}$&$(3.70\pm1.35)\times10^{-4}$ &$(0.68\pm0.46)\times10^{-4}$&$-$ \\     \hline
		$D^0\rightarrow K^0\eta$  &$(0.94\pm0.45)\times10^{-4}$&$(0.28\pm0.10)\times10^{-4}$ &$(0.96\pm0.32)\times10^{-4}$&$-$ \\   \hline
		$D^0\rightarrow K^0\eta'$ &$(8.02\pm3.32)\times10^{-4}$ &$(0.59\pm0.08)\times10^{-4}$&$(9.22\pm1.61)\times10^{-4}$&$-$ \\     \hline
		$D^+\rightarrow K^0\pi^+$ &$(3.27\pm1.86)\times10^{-4}$&$(1.19\pm0.55)\times10^{-3}$ &$(3.51\pm2.11)\times10^{-4}$&$-$ \\     \hline
		$D^+\rightarrow K^+\pi^0$ &$(3.07\pm1.02)\times10^{-4}$&$(2.15\pm1.17)\times10^{-4}$ &$(3.27\pm1.39)\times10^{-4}$&$(1.83\pm0.26)\times10^{-4}$ \\   \hline
		$D^+\rightarrow K^+\eta$  &$(0.98\pm0.26)\times10^{-4}$ &$(1.04\pm0.23)\times10^{-4}$ &$(0.89\pm0.27)\times10^{-4}$&$(1.08\pm0.17)\times10^{-4}$ \\   \hline
		$D^+\rightarrow K^+\eta'$ &$(1.40\pm0.39)\times10^{-4}$ &$(1.82\pm0.18)\times10^{-4}$&$(1.35\pm0.39)\times10^{-4}$&$(1.76\pm0.22\times10^{-4})$ \\   \hline
		$D_s^+\rightarrow K^+K^0$   &$(7.84\pm2.31)\times10^{-4}$ &$(0.68\pm0.09)\times10^{-4}$&$(0.72\pm0.44)\times10^{-4}$&$-$ \\   \hline
		\hline 
	\end{tabular}
	\label{table:BRvalDCS} 
\end{table}

\section{Conclusions}\label{Conclusions}
For several decades, various ratios of decay rates of many of the $D\rightarrow PP$ modes remained to be a puzzle as these were expected to be one in the $SU(3)$ limit, but the measured values exhibited 
large deviations from unity.

We have evaluated the bare amplitudes of all the $D\rightarrow PP$ modes using factorization, however, we add non-factorizable corrections, weak annihilation and exchange contributions as parameters, 
and in the hadron matrix elements, the $q^2$ dependence of the form factors involved are evaluated using the z-expansion method and finally,  resonant final state interaction effects are incorporated. 
The parameters of the $K$ matrix coupling the various channels are defined using the measured decay widths of the resonances (where available) and those unobserved, are left as parameters to be fitted 
from all the measured 28 $D\rightarrow PP$ branching ratios. 
Our best fit has a $\chi^2$/degree of freedom of 2.25, which is an improvement over the previous results in Refs.~\cite{Li:2012cfa,Cheng:2010ry}.                                      
\begin{itemize}
\item We are able to get reasonable fits to almost all the observed branching ratios. In particular, our branching fractions for $D\rightarrow KK$, $D\rightarrow\pi\pi$ modes that have been a 
long standing puzzle are in agreement with the corresponding measured values.
\item We have  evaluated the cosine of the strong phase difference between the unitarized amplitudes for $D^0\rightarrow K^-\pi^+$ and $D^0\rightarrow K^+\pi^-$ and obtain, $\cos\delta_{K\pi}=0.94\pm0.027$. 
This result is consistent with the  recently measured  BESIII result, $\cos\delta_{K\pi}= 1.02\pm0.11\pm0.06\pm0.01$ ~\cite{Ablikim:2014gvw}.
\item The mode $D^0\rightarrow K^0\bar{K^0}$ does not have any tree or colour suppressed contributions, but can come only from W-exchange. In fact, there are two exchange contributions, one appearing with 
a $d\bar{d}$ and the other with an $s\bar{s}$, which under exact $SU(3)$ symmetry would cancel  each other, resulting in a null amplitude. However, since our parameters for these two contributions are 
distinct, 
our bare amplitude for this mode is small but non-vanishing. There have been speculations ~\cite{Cheng:2002wu, Zenczykowski:1996xd} that this mode can arise just from final state interactions, even in the 
absence of a weak exchange contribution. However, from Table~\ref{table:BRvalSCS} it is clear that without the exchange contribution we are unable to generate a large enough rate: both final state interaction
and the exchange contribution are necessary for consistency with the measured branching fraction.                                                                                                
\item We have also evaluated the four ratios of amplitudes that had been specified in a recent paper~\cite{Gronau:2015rda}. In SU(3) limit these are all expected to be unity. Our theoretical estimates for 
these ratios are given below:
\begin{eqnarray*}
 R_1&\equiv&\frac{|A(D^0\rightarrow K^+\pi^-)|}{|A(D^0\rightarrow\pi^+K^-|\tan^2\theta_c}=1.27\pm0.32,\\
 R_2&\equiv&\frac{|A(D^0\rightarrow K^+K^-)|}{|A(D^0\rightarrow\pi^+\pi^-|}=1.27\pm0.42,\\
 R_3&\equiv&\frac{|A(D^0\rightarrow K^+K^-)|+|A(D^0\rightarrow\pi^+\pi^-|}{|A(D^0\rightarrow\pi^+K^-|\tan\theta_c+|A(D^0\rightarrow K^+\pi^-)|\tan^{-1}\theta_c}=1.19\pm0.28,\\
 R_4&\equiv&\sqrt{\frac{|A(D^0\rightarrow K^+K^-)||A(D^0\rightarrow\pi^+\pi^-|}{|A(D^0\rightarrow\pi^+K^-||A(D^0\rightarrow K^+\pi^-)|}}=1.19\pm0.26.
\end{eqnarray*}
Furthermore, the following combination of these ratios is expected to be vanishing up to $4^{\text{th}}$ order in U-spin breaking,
\begin{equation*}
 \Delta R\equiv R_3 - R_4 +\frac{1}{8}\left[\left(\sqrt{2R_1-1}-1\right)^2-\left(\sqrt{2R_2-1}-1\right)^2\right].
\end{equation*}
Using our unitarized amplitudes, we find the central value of $\Delta R$ to be indeed very tiny, however, with a large error.
\begin{equation*}
 \Delta R=-0.000013\pm0.006. 
\end{equation*}
Our theoretical errors (in form factors, K-matrix parameters, etc.) are propagated to evaluate the errors in the ratios $R_1$, $R_2$, $R_3$, $R_4$ and finally in $\Delta R$.

\item We would like to mention that in many modes involving $\eta$ and $\eta'$, we have additional terms in our amplitude due to our distinction of the different form factors, compared to, for eg., those 
that appear in Ref.~\cite{Cheng:2010ry}. A naive look at the colour suppressed diagrams for $D\to \pi^0\eta(\eta')$ will indicate that the contributions from the case where the spectator is part of the $\pi^0$, and that, where it constitutes the $\eta(\eta')$ must cancel. However, in terms of the specific decay constants and form factors, one is proportional to $-f_{\pi}F^{D\eta_q}_0(m_{\pi}^2)$ while the other is proportional to $f_{\eta_q}F^{D\pi}_0(m_{\eta}^2)$, which are unequal and hence must survive. 
\item While the Particle Data Group~\cite{Agashe:2014kda} does not include a world average for $\Gamma(f_0(1710)\rightarrow K\bar{K})$ but it does list  two values  for the ratio $\frac{\Gamma(f_0(1710)\rightarrow K\bar{K})}{\Gamma_{total}}$: $0.36\pm0.12$ (Ref.~\cite{PhysRevLett.101.252002}) and $0.38^{+0.09}_{-0.19}$ (Ref.~\cite{Longacre:1986fh});
our fit value of $g_{pe}(\equiv\Gamma(f_0(1710)\rightarrow K\bar{K}))$ corresponds to 0.35 for the branching ratio, which is consistent with these values.
\item Our theoretical errors are rather large and could be reduced in future with more precise form factors available either from measurement of semileptonic $D$, $D_s$ modes at BES III, where if even the 
lepton mass could be incorporated (eg. by looking at modes with muon in the final state), then the $q^2$ dependent $F_0$ could be known, or with improved lattice studies, specially for $D\rightarrow \eta'$, $D_s\rightarrow K$ and $D_s\rightarrow\eta,\eta'$.                                               
\item Accurate measurements of the decay widths  of the resonances (used for the final state interactions) to many of the coupled channels can reduce the theoretical uncertainties 
and possibly allow for better fits to data. For example our fits seem to indicate a rather large value for $h_{pe}$ or the width of $a_0(1450)\rightarrow\pi^0\eta$. This seems to result in larger branching 
fractions for many of the isospin one modes. Future measurement of this width can help reduce this uncertainty and perhaps result in better fits to data for these modes. 
\item Out of the 28 observed PP modes, we are unable to fit 7 of the modes well. Many of these modes involve $\eta$ or $\eta'$ in the final state. Including a gluonium component in the $\eta$, $\eta'$ states may possibly be one way of improving these fits. This, along with improved form factor measurements, observation of decay rates of the resonances (playing a role in final state interactions) to these decay modes, as mentioned in the last two points above, could go a long way in improving our fits. One glaring misfit is the mode $D^+\to K^+\bar{K^0}$. This mode does couple to $\pi^+\eta, \pi^+\eta'$ modes and hence, may possibly improve, along with the improvements in those modes.  Finally, perhaps a more sophisticated statistical analysis may also play a role in improving our results, which we hope to carry out in future~\cite{Biswas2}. 
\end{itemize}

\appendix 
 
\section{The Wilson Coefficients}  
 
The Wilson coefficients used in the evaluation of all the bare amplitudes have been calculated  at the final state hadronic scale. This allows for an additional $SU(3)$ breaking effect and they have 
been incorporated using the procedure outlined in Ref.~\cite{Li:2012cfa}. The Wilson coefficients at lower scale are calculated in the Ref.~\cite{Fajfer:2002gp}.  The essential steps are following.     

 \begin{itemize}
 
 \item 
In the first step, the Wilson coefficients $C_i (m_W)$ at weak scale are calculated by requiring the equality of the effective theory with five active
flavors $q = u, d, s, c, b$ onto the full theory. 

 \item
Next, the coefficients undergo the evolution from the scale $m_W$ to $\mu$  through the equation
  
  \begin{equation}
   C (\mu) = U_5 (\mu,m_W)  C (m_W).
  \end{equation}
  
  \item
  In the next step, coefficients are calculated at the scale of $b$ quark 
  
  \begin{equation}
  C (m_b) \rightarrow Z (m_b).
  \end{equation}
  
  \item
Now, the Wilson coefficients can be evaluated at required scale ($\mu_c$ or $\mu_{hadron}$) through the equation

  \begin{equation}
 C (\mu)=U_4(\mu,m_b)  Z (m_b).
  \end{equation}  
 In the above steps, $U_5$ and $U_4$ are the $2\times  2$  and $7\times  7$ evolution matrices for five and four active flavors respectively.  The  $Z(m_b)$ is given in the Eqs.$A.7$ to $A.10$ of the 
Ref.~\cite{Fajfer:2002gp}.
  \end{itemize}
 The explicit  expressions of the Wilson coefficients obtained after following the above steps are given in the Ref.~\cite{Li:2012cfa}.  They are:

            \begin{eqnarray*}
            	C_1(\mu)=&-0.2334\alpha^{1.444}+0.0459\alpha^{0.7778}+1.313\alpha^{0.4444}-0.3041\alpha^{-0.2222},&\\
            	C_2(\mu)=&0.2334\alpha^{1.444}+0.0459\alpha^{0.7778}-1.313\alpha^{0.4444}+0.3041\alpha^{-0.2222}&.
            \end{eqnarray*}
            
            in terms of the running coupling constant $\alpha$:
            \begin{equation*}
            \alpha=\alpha_s(\mu)=\frac{4\pi}{\beta_0ln(\mu^2/\Lambda_{\overline{MS}}^2)}\left[1-\frac{\beta_1lnln(\mu^2/\Lambda_{\overline{MS}}^2)}{\beta_0^2ln(\mu^2/\Lambda_{\overline{MS}}^2)}\right],
            \end{equation*}
            
            with the coefficients
            \begin{eqnarray*}
            	\beta_0=\frac{33-2f}{3}\;\;\;,\;\;\; \beta_1=102-\frac{38}{3}N_f
            \end{eqnarray*}
            
 We take active flavour number $N_f=3$, and the QCD scale $\Lambda_{\overline{MS}}=\Lambda_{\overline{MS}}=375$ MeV. Again, note that the scale dependent strong coupling constant $\alpha_s(\mu)$ is evaluated 
at the final state hadronic scales for each individual decay, to take care of the SU(3) breaking. 
                        
\section{Series Expansion Method for Form Factors}  
Using the analytic properties of $F(q^2)$, a transformation of variable is made which maps the cut on the $q^2$ plane onto a unit circle $|z|<1$, where
\begin{eqnarray*}
 z(t_, t_0)=\frac{\sqrt{t_+ - t}-\sqrt{t_+ - t_0}}{\sqrt{t_+ - t}+\sqrt{t_+ - t_0}} \;\;\;,\;\;\;t=q^2,
\end{eqnarray*}
 where,
 \begin{equation*}
  t_{\pm} = (m_D \pm m_{P_1})^2 \text{ and }
 t_0 = t_+ \left(1 - \left(1 -\frac{t_-}{t_+}\right)^{1/2}\right).
 \end{equation*}
 This transformation allows the form factors to be given by an expansion about $q^2=t_0$, given as
\begin{equation*}
F(t) = \frac{1}{P(t)\phi(t,t_0)}\sum_{\mathrm k=0}^{\inf} a_k(t_0)z(t,t_0)^k,~
\end{equation*}
 given also as Eq.(~\ref{z-exp}) in the text.
 The function $P(t)$ in the above is 1 for $D\rightarrow\pi$ form factors. For $D_s\rightarrow\eta$ and $D\rightarrow K$ form factors, $P(t)=z(t,M_{D^*_{s0}}^2)$ (where $M_{D^*_{s0}}$) is the nearest $0^+$ 
resonance mass).                                                               
The outer function $\phi$ is given by \cite{Arnesen:2005ez}
\begin{equation*}	
	\phi(t,t_0)=\sqrt{\frac{3 t_+ t_-}{32 \pi\chi_0}} \left(\frac{z(q^2,0)}{-q^2}\right)^{2}\left(\frac{z(q^2, t_0)}{t_0 - q^2}\right)^{-1/2}
\left(\frac{z(q^2, t_-)}{t_- - q^2}\right)^{-1/4}\frac{\sqrt{t_+ - q^2}}{(t_+ - t_0)^{1/4}},
\end{equation*}	
         
where $\chi_0$ has been calculated \cite{Arnesen:2005ez} using OPE and is given by:  
\begin{equation*}
\chi_0 = \frac{1+0.751 \alpha_s (m_c)}{8 \pi^2}. 
\end{equation*}  
  For simplicity, we ignore condensate contribution which is of the order  $\mathcal{O} (m_c^{-3})$ and  $\mathcal{O} (m_c^{-4})$  The strong coupling at charm scale is computed with the package RunDec 
\cite{Chetyrkin:2000yt}.
            
\acknowledgements 

The authors thank L.~Gibbons, R.~Hill and J.~Libby for correspondence related to form factors. NS thanks H.-Y.~Cheng for discussions at the early stage of this work and A.~Paul, S.~Pakvasa and N.~G.~Deshpande for comments. She also thanks T.~E.~Browder for information regarding the BESIII strong phase measurement. 
AB thanks S.~Patra for suggestions regarding the numerical analysis. GA thanks A.~Pich for discussions. GA's work has been supported in part by the Spanish Government and ERDF funds from the EU Commission [Grants No. FPA2011-23778, FPA2014-53631-C2-1-P No. CSD2007-00042 (Consolider Project CPAN)].

\end{document}